\documentclass[12pt]{article}
\newwrite\ffile\global\newcount\figno \global\figno=1

\def\writedef#1{}

\input epsf
\def\figin{\epsfcheck\figin}\def\figins{\epsfcheck\figins}
\def\epsfcheck{\ifx\epsfbox\UnDeFiNeD
\message{(NO epsf.tex, FIGURES WILL BE IGNORED)}
\gdef\figin##1{\vskip2in}\gdef\figins##1{\hskip.5in}
\else\message{(FIGURES WILL BE INCLUDED)}%
\gdef\figin##1{##1}\gdef\figins##1{##1}\fi}

\def\figinsert{}
\def\ifig#1#2#3{\xdef#1{fig.~\the\figno}
\writedef{#1\leftbracket fig.\noexpand~\the\figno}%
\figinsert\figin{\centerline{#3}}\medskip\centerline{\vbox{\baselineskip12pt
\advance\hsize by -1truein\center\footnotesize{  Fig.~\the\figno.} #2}}
\bigskip\endinsert\global\advance\figno by1}
\def\endinsert{}

\begin{document}
\baselineskip 18pt
\newcommand{\Tr}{\mbox{Tr\,}}
\newcommand{\beq}{\begin{equation}}
\newcommand{\eeq}{\end{equation}}
\newcommand{\bea}{\begin{eqnarray}}
\newcommand{\eea}[1]{\label{#1}\end{eqnarray}}
\renewcommand{\Re}{\mbox{Re}\,}
\renewcommand{\Im}{\mbox{Im}\,}
\begin{titlepage}

\begin{picture}(0,0)(0,0)
\put(350,0){SHEP-01-31}
\end{picture}
 
\begin{center}
\hfill
\vskip .4in
{\large\bf Field Theory Operator Encoding in $\mathcal{N}=2$ Geometries}
\end{center}
\vskip .4in
\begin{center}
{\large James Babington and Nick Evans}
\footnotetext{e-mail: jrb4@hep.phys.soton.ac.uk, 
n.evans@hep.phys.soton.ac.uk }
\vskip .1in
{\em Department of Physics, Southampton University, Southampton,
S017 1BJ, UK}

\end{center}
\vskip .4in
\begin{center} {\bf ABSTRACT} \end{center}
\begin{quotation}We investigate supergravity solutions describing 
D5 branes wrapped on a two cycle
which are dual to $\mathcal{N}=2$ super Yang Mills theory. 
Brane probing these solutions allows the moduli space of the field theory 
to be identified. There are a unique set of coordinates in which the field 
theory on the probe takes an $\mathcal{N}=2$ form and in these coordinates 
the running coupling of the gauge theory may be identified. We show that 
the geometry, when restricted to the moduli space, takes a very simple
form involving only two functions. One is the running gauge coupling whilst the
other parametrizes the scalar operators of the field theory. The D5 brane 
distributions, for the full set of solutions in the literature,
can be determined by assuming the field theory's form for 
the running coupling as a function of scalar vevs. We show that the resulting
distributions also correctly 
reproduce the scalar operators parametrized elsewhere
providing the first non-trivial consistency check on the duality.
\noindent 

\end{quotation}
\vfill
\end{titlepage}
\eject
\noindent

\section{Introduction}
The $AdS/CFT$ correspondence~\cite{conjecture} has enabled the study of
a wide class of strongly coupled gauge theory phenomena from a novel 
supergravity point of view. Much work has been done to expand these ideas 
to a wider class of such dualities. One approach has been to start 
with the original conjecture, describing ${\cal N} = 4$ super Yang Mills (SYM)
theory and introduce deformations by turning on supergravity fields
~\cite{5ddeform, freed2, N=2star}. In parallel, new dualities 
have been obtained \cite{construct}
by considering the near horizon limits of supergravity backgrounds around 
various D brane configurations with less supersymmetric field theories on 
their world volumes. In both cases one is limited by the complexities
of supergravity calculations and typically solutions can only be found
for very particular configurations describing the field theory at restricted
points on moduli space. Ideally one would like to be able to deduce from
these solutions how the supergravity background encodes the dual field
theory operators and couplings so more complete solutions may be found.
In \cite{jambab} 
we showed that it is possible to (retrospectively) obtain the full
multi-centre D3 brane solutions \cite{brr,freed2} that describe ${\cal N} = 4$
SYM on moduli space from deformed 5d supergravity solutions lifted to
10d. In this paper we apply these ideas to geometries describing ${\cal N} = 2$
SYM \cite{martelli, zaff} 
and show how the field theory operators are encoded in the background.  
 
We will study the $\mathcal{N}$=2 geometries realized in
\cite{martelli, zaff} by considering D5 branes wrapped on $S^2$.
The geometries describe restricted points on the field theory 
moduli space and were obtained from 7d gauged supergravity solutions
which were then lifted to 10d solutions. Those
authors have already performed D5 brane probes on the resulting 
solutions and found the unique coordinates appropriate for
the duality by insisting that the U(1) field theory on the probe's
world volume takes $\mathcal{N}$=2 form. They showed that the solutions
have a running coupling, as anticipated by the field theory,
which diverges at an enhan\c con locus \cite{enhan}
for some of the geometries (in this sense the solutions are
similar to the ${\cal N}=2^*$ geometries discussed in \cite{N=2star,jambab}). 
We will pick up in the ${\cal N}=2$ geometries  at this point
and show that the geometry when restricted to the moduli space takes
a very simple form with only two free functions. One of these functions 
is simply the gauge coupling of the probe brane theory whilst the other
is essentially the dilaton. The duality would lead one to expect that
this remaining function should parametrize the scalar vevs of the field
theory vacuum. We solve for the function in the coordinates appropriate
to the duality and interpret the coefficients it contains in this 
light from their symmetry properties.

In \cite{zaff} the $\mathcal{N}$=2 field theory expectation
for the form of the running coupling as a function of scalar vevs 
was imposed on the simplist supergravity solutions to determine the 
D5 brane distribution function. We extend this analysis to provide the
distribution function for the full set of solutions. Using this 
distribution function we can then predict the scalar operators in the field 
theory. They match the operators parametrized in the dilaton piece of
the background providing the first non-trivial check that the 
supergravity solution is governed by the $\mathcal{N}$=2 field theory 
dynamics. 

In this paper we restrict ourselves to the sub-space of the geometry
which describes the moduli space of the probe theory 
rather than the full solution so this work is only a step towards 
understanding how to expand the known set of solutions. Nevertheless
we consider it an important step in that direction.

\section{The $\mathcal{N}$=2 Supergravity Solutions}

\subsection{The 7d Background}

We collect here the relevant facts which we will need from 
the supergravity solutions obtained in 
\cite{martelli, zaff}. At the level of 7d supergravity 
the metric in the string frame takes the form

\begin{equation}
ds^2_7=dx^2_4+R^2e^{2h}d\Omega^2_2+d\rho^2,
\end{equation}
Here $x_4$ are the four directions parallel to the unwrapped part of the brane,
$R$ is the radius of the $S^2$ parametrized by $\Omega_2$ that we wrap the 
remaining brane dimensions on; also $h$ measures supergravity deformations 
and $\rho$ is the radial coordinate. By looking at the supersymmetry 
variations of
the fermionic fields, and setting these to zero, first
order equations can be obtained for the bosonic fields describing 
$\mathcal{N}$=2 preserving deformations. The fields considered in
\cite{martelli, zaff} are the scalars
$(\lambda_1,\lambda_2,\tilde{\lambda_2})$ which enter the metric through

\begin{equation} h= g-f 
\end{equation}
where the supergravity equations of motion are
\begin{eqnarray}
f' &=& -({\lambda_{1}}' +{{{\lambda_{2}}'+{\tilde\lambda_2}'}\over 2}), \\
g' &=& -({\lambda_{1}}' +{{\lambda_{2}}'+{\tilde\lambda_2}'\over 2}) +{1 \over 2\Lambda}e^{f-2g-2\lambda_{1}}, \\
{\lambda_{2}}'+2{\tilde\lambda_{2}}'+2{\lambda_{1}}' &=& -{1\over \Lambda}e^{f+2\tilde\lambda_{2}}, \\
{2\lambda_{2}}'+{\tilde\lambda_{2}}'+2{\lambda_{1}}' &=& -{1\over  \Lambda}e^{f+2\lambda_{2}}, \\
3{\lambda_{1}}'+{\lambda_{2}}'+{\tilde\lambda_{2}}' &=& -{1\over \Lambda}e^{f+2\lambda_{1}}+{1\over 2 \Lambda}e^{f-2g-2\lambda_{1}},
\label{first2}
\end{eqnarray}
a prime indicates differentiation with respect to the radial coordinate 
$\rho$, and $\Lambda$ is a dimensionful constant, which will play the 
role of the strong coupling scale in the dual field theory.

Note that $\lambda_2$ and $\tilde{\lambda_2}$ enter the equations in a
symmetrical way. This full set of equations was studied in \cite{zaff} 
whilst in \cite{martelli} only the case $\lambda_2=\tilde{\lambda_2}$
was considered. Defining the
dimensionless radial coordinate $u$, and making the change from
$\rho\rightarrow u$;
\begin{eqnarray}
u &\equiv &e^{2h}, \\ {\Lambda \over H(u)}\equiv \frac{d\rho}{du} &\equiv &
\Lambda
e^{\lambda_1-1/2(\lambda_2+\tilde{\lambda_2})} \label{eq:coch}
\end{eqnarray}

We can extract solutions for a number of quantities we shall use later.
It is straightforward to  show,
from the field equations above, that
\begin{equation}
e^{-\lambda_2+\tilde{\lambda_2}}=\frac{e^{2u}+b^2}{e^{2u}-b^2}. \label{eq:sol1}
\end{equation}
where $b$ is an integration constant, and
\begin{equation} 
e^{-4 \lambda_1 - 2 \lambda_2 - 4 \tilde{\lambda}_2 } = c e^{-2 u}
\left( e^{2u} - 
b^2 \right)^2 \label{eq:sol2}
\end{equation}
where $c$ is an integration constant, which only appears as an overall
factor in the analysis below so we set it to one.

The final solution we will need is an expression for the function 
$H$ which  satisfies the equation

\begin{equation}
\frac{dH^2}{du}+H^2\left(\frac{1}{u}+2\left(\frac{e^{4u}+b^4}{e^{4u}-b^4}\right)\right)=2,
\end{equation}
We can explicitly solve this finding
\begin{equation}
H^2(u)=\left[\frac{e^{4u}+b^4}{e^{4u}-b^4}-\frac{1}{2u}+
\frac{2K}{u \Lambda^2}
\left(\frac{e^{2u}}{e^{4u}-b^4}\right)\right].\label{eq:sol3}
\end{equation}
where $K$ is again an integration constant (we have scaled $K$ by 
$\Lambda$ so it has the correct dimensions below). Note that when 
$\lambda_2 = \tilde{\lambda}_2$ as investigated in \cite{martelli} 
$b=0$ and in this limit we recover their solution
\begin{equation}
\lim_{b \rightarrow 0} H^2(u)=1-{1\over 2u} + {2K e^{-2u} \over u \Lambda^2}. 
\end{equation}

\subsection{The 10d Background}

To brane probe these solutions they must be lifted to 10d. 
The lift was performed in~\cite{zaff} and we have the
string frame solution
\begin{eqnarray}
 ds^2 &=& ds^2_7+ e^{2\lambda_1+\lambda_2+\tilde{\lambda_2}}
\Delta^{-1}
[e^{-2\lambda_1}[d\mu_1^2+d\mu_2^2+\cos^2\theta(\mu_1^2+\mu_2^2)
d\phi^2 \nonumber \\ & &
-2\cos\theta(\mu_1d\mu_2 +\mu_2d\mu_1)d\phi]+e^{-2\lambda_2}d\mu_3^2+ e^{-2\tilde{\lambda}_2}d\mu_4^2],
\end{eqnarray}
with

\begin{eqnarray}
e^{2\Phi} &=&
e^{6\lambda_1+3\lambda_2+3\tilde{\lambda_2}}\Delta^{-1},
\\
 \Delta &=& e^{2\lambda_1}(\mu_1^2+\mu_2^2)+e^{2\lambda_2}\mu_3^2+
 e^{2\tilde{\lambda_2}}\mu_4^2.
\end{eqnarray}
The additional $S^3$ parameterization is given by the coordinates $\mu_i$,
such that $\sum_{i=1}^4\mu_i^2=1$. These can be written in terms of
the usual three angles, 
\begin{eqnarray}
(\mu_1,\mu_2)&=&
\sin\psi (\cos\phi_1,\sin\phi_1) \\
(\mu_3,\mu_4)&=&
\cos\psi (\cos\phi_2,\sin\phi_2)
\end{eqnarray}

These solutions in fact describe the near horizon geometry of NS5 branes
wrapped on $S^2$. To convert from a NS$5$ solution to a D5 solution
one performs an  $S$-dual transformations~\cite{zaff}

\begin{eqnarray}
\Phi_D &=&-\Phi, \nonumber \\
ds^2_D &=&e^{\Phi_D}ds^2_{NS}.
\end{eqnarray}

There is also a 6 form potential for which the D5 branes are sources.
The full expression is not given in \cite{zaff} but the components in
the D5 world volume, when $\mu_1=\mu_2=0$, 
relevant to the brane probe analysis below is given by

\beq
C_6 = R^2 e^{2 \Phi_D}u  ~ dx_4 \wedge d\Omega_2 
\eeq

At this point then, we have the string frame $D5$ solution, 
which we can now proceeed to brane probe. The full details of these solutions 
and lifts can be found in~\cite{zaff}.

\section{The 2d Moduli Space and Distributions}

Each geometry, corresponding to a solution of (\ref{first2}), 
is expected to be dual
to the ${\cal N}=2$ SYM theory at a point on its 2d moduli space. Since the 
${\cal N}=2$ SYM theory has a moduli space, each of these solutions should
then display a 2d space in which a probe D5 brane sees a flat potential.
This corresponds to the theory knowing that any individual scalar vev 
may be changed at will on the moduli space whilst keeping a vacuum. At large
N the changing of such a single vev, or position of a D5, will leave the
geometry unchanged. In \cite{zaff} it was shown that such a 2d moduli
space does indeed exist for all of these solutions.

For a single wrapped $D5$-brane, we have the low energy effective 
Born-Infeld action

\begin{equation}
S_{probe}=-T_5\int d^6\xi e^{-\Phi_D}\sqrt{-\det (G_{ab}+F_{ab})}+Q_5\int
C_6.
\end{equation}
where $\xi$ are coordinates on the brane, $G_{ab}$ is the pullback of the $10d$ spacetime metric, $F_{ab}$ is the surface gauge field strength and 
$T_5=Q_5 g_s^{-1}$.

As in~\cite{zaff}, setting $\mu_1=\mu_2=0$, and substituting in
the background we find the gauge
potential cancelling against the leading term  from the 
expansion of the square root.
Thus the $(u, \phi_2)$ plane is the moduli space.  
From henceforth we restrict 
ourselves to this moduli space since only on this space can we use field
theory intuition in the probe world volume theory to find the correct
coordinates in which to interpret the duality.

If we allow the probe brane to move slowly on the moduli space and also 
allow small gauge fields on its surface we can find the leading
kinetic terms in the probe world volume theory. 
Passing to a new radial coordinate
$u=\ln({z/\Lambda})$, we may write the kinetic piece in the form

\begin{equation} \label{pro}
S_{probe} =-T_5 R^2 \int d\Omega_2 d^4x    {\cal L}
\end{equation}
where

\begin{eqnarray}
{\cal L} &=& 
\frac{\ln{(z/\Lambda})}{(z/\Lambda)^2}
e^{-4\lambda_1-2\lambda_2-4\tilde{\lambda_2}}[\cos^2\phi_2
+e^{-2\lambda_2+2\tilde{\lambda_2}}\sin^2\phi_2](\dot{z}^2
+z^2\dot{\phi_2}^2)  \nonumber \\  
& & \hspace{4cm}+ {1 \over 4} \ln (z / \Lambda)  F^{\mu \nu} F_{\mu \nu}.
\end{eqnarray}

Note that it was the choice  of coordinate transformation in (\ref{eq:coch}), 
which allowed us to
factor out the scalar terms between the coordinate $u$ and the
angular pieces $\mu_1,\mu_2$.

Using the field equations in (\ref{eq:sol1})(\ref{eq:sol2}) we can evaluate this to be

\begin{eqnarray}
{\cal L} &=& \frac{\ln{(z/\Lambda})}{(z/\Lambda)^4}
[(z^2/\Lambda^2-b^2)^2\cos^2\phi_2 
 +(z^2/\Lambda^2+b^2)^2\sin^2\phi_2](\dot{z}^2+z^2\dot{\phi}_2^2) \nonumber \\
& &\hspace{4cm}+ {1 \over 4} \ln (z / \Lambda)  F^{\mu \nu} F_{\mu \nu},
\end{eqnarray}
which can be written in terms of the complex coordinate
$Z=ze^{i\phi_2}$  as
\begin{equation} \label{zoom}
{\cal L} =  \ln \left({|Z| \over \Lambda} \right)
\left[\left(1-\frac{b^2 \Lambda^2}{Z^2}\right)\left(1-\frac{b^2 \Lambda^2}
{\bar{Z}^2}\right)\right]
|\dot{Z}|^2
+ {1 \over 4} \ln \left({|Z| \over \Lambda}\right)  F^{\mu \nu} F_{\mu \nu}.
\end{equation}

This form for the solution does not display the explicit ${\cal N}=2$ 
form of the field theory. To find such a form  we need to pass to a 
new set of coordinates, $W$, such that the scalar and gauge kinetic pieces 
appear with the right normalization

\begin{equation}
{\cal L} = {1 \over g^2_{YM}(W)} \left( |\dot{W}|^2 + {1 \over 4} 
F^{\mu \nu} F_{\mu \nu} \right) \label{eq:nice}
\end{equation}

The appropriate Jacobian and hence  the appropriate holomorphic 
change of variables \cite{zaff} may be seen from (\ref{zoom})

\begin{equation}
W=Z+b^2\Lambda^2/Z. \label{eq:W}
\end{equation}
The gauge coupling now reads
\begin{equation}
{1 \over g_{YM}^2(W)} = \ln \left({|Z(W)| \over \Lambda} \right)
 =\cosh^{-1} (\frac{W}{2b\Lambda})+\ln b. \label{eq:couple}
\end{equation}

\section{The Geometry of the Moduli Space and D5 Distributions}

We have identified the unique set of coordinates on the moduli space, $W$,
where the field theory duality is manifest. For the brane probe
to leave the world volume theory (\ref{eq:nice}) the geometry 
in the sub-space corresponding to the moduli space must 
take the form  

\begin{equation}
ds_D^2=e^{\Phi_D}[dx_4^2+ {1 \over g^2_{YM}(W)} 
R^2 d\Omega_2^2+{e^{-2\Phi_D} \over \Lambda^4} dWd\overline{W}]
\end{equation}

\begin{equation}
C_6 = R^2 e^{2 \Phi_D} { 1 \over g^2_{YM}(W)} dx_4 \wedge d \Omega_2 
\end{equation}

Note we have introduced the factors of $\Lambda$ in the metric component 
$G_{ww}$ in order to make the four dimensional lagrangian in (\ref{pro})
dimensionally correct. 

The background is described by two functions. One we have identified
as the gauge coupling whilst the other, $e^{\Phi_D}$, 
remains to be interpreted.
We may find an explicit expression for the dilaton from the metric element
$G_{ww}$ which is $G_{zz}$ times the Jacobian for the transformation. 
We find in the $z$ coordinates

\begin{equation}
e^{2\Phi_D}=\frac{H^2 \Lambda^2}{z^2}[(z^2/\Lambda^2-b^2)^2\cos^2\phi_2
 +(z^2/\Lambda^2+b^2)^2\sin^2\phi_2].
\end{equation}
and our solution for $H$ from (\ref{eq:sol3}) in these coordinates is
\begin{equation}
H^2(z)=\left[\frac{z^4/\Lambda^4+b^4}{z^4/\Lambda^4-b^4}-{1
\over \ln (z/ \Lambda)}+
{2 K  \over \ln(z/ \Lambda)}\left(\frac{z^2}
{z^4-b^4\Lambda^4}\right)\right]. \label{eq:H}
\end{equation}

To attempt to interpret this function in terms of the field theory 
we must translate it to the coordinates appropriate to the duality, $W$.
In fact, simply following through the holomorphic change of
coordinates, we find 
\begin{equation}
e^{2\Phi_D}=\frac{W\bar{W}}{\Lambda^2}G^{1/2}\bar{G}^{1/2}\left[\frac{1+
G^{1/2}\bar{G}^{1/2}}{G^{1/2}+\bar{G}^{1/2}}-{1 
\over g^2_{YM}(W)}+
{2 K 
\over g^2_{YM}(W)}\left(\frac{2}{W\bar{W}(G^{1/2}
+\bar{G}^{1/2})}\right)\right], \label{ans}
\end{equation}
where
\begin{equation} \label{g}
G=1-\frac{4b^2\Lambda^2}{W^2}.
\end{equation}

Let us now consider the anatomy of the solution in both the 
$Z$ and $W$ coordinates. Firstly looking in the $Z$ coordinates the 
solution has no $\phi_2$ dependence so the D5 brane 
distribution must be symmetric in the $Z$ plane. 
As can be seen from (\ref{eq:H}) there is always a singularity in the metric at
$z=b$. For large $K$ though there can be a singularity at larger $z$. In
fact we can trade the parameter $K$ for the radius of the singularity $z_0$

\begin{equation} \label{k}
K=\frac{z_0^4-b^4}{4z_0^2}- \ln (z_0/\Lambda) 
\left(\frac{z_0^4+b^4 \Lambda^4}{2z_0^2}\right).
\end{equation}
The function $\ln (z_0/\Lambda)$ when translated to the physical $W$ 
coordinates has the simple interpretation of 
$1/g^2_{YM}$ evaluated at the position of the 
singularity and we will write it henceforth as $1/g^2_{YM}({\rm sing})$.

We shall interpret the singularity as indicating the position of the D5
branes. Note that any given solution only describes the space 
$z_0 < z < \infty$. A probe is therefore restricted to this space
and thus only for $z_0=b=1$ can it reach the enhan\c con locus (a circle here)
where the coupling diverges. That distribution
must correspond to a singular point on the field theory moduli space. 

For large $z_0$ we may neglect $b$ and the distribution is essentially 
a circle in the physical $W$ coordinates (since $Z \simeq W$). As $z_0$
reduces, the coordinate transformation to $W$ in (\ref{eq:W}) distorts the circle
by squashing it in the imaginary $W$ direction.  When $z_0 = b$ the
singularity lies on the real line between $w = \pm 2b$. 

We would like to find the explicit distribution function for the D5 branes
$\sigma_w(W)$ in the physical coordinates.
We can attempt to do this using the supergravity expression for $g^2_{YM}$
and the form of the ${\cal N}=2$ field theory prediction for the coupling
as a function of scalar vevs. We would expect

\begin{equation}
{4 \pi \over g^2_{YM}(W) } = 
{1 \over \pi} \int \sigma_w(A) \ln \left({A-W \over \Lambda} \right) 
dA d\bar{A} \label{ft}
\end{equation}

However, given the complicated shape of the distribution in $W$ it is
easier to transform this equation to the $Z$ coordinates where we know we have
spherical symmetry

\begin{equation}
{4 \pi \over g^2_{YM}(W) } = {1 \over \pi} \int \sigma_z(Z) 
\ln \left({Z + {b^2 \Lambda^2 \over Z} -W \over \Lambda} \right) dZ d \bar{Z}
\end{equation}

A degree of guess work is required to find the appropriate $\sigma_z$ that
reproduces the coupling in (\ref{eq:couple}). In fact the simple guess that the distribution
is just a ring at $z=z_0$ reproduces the supergravity result. Thus

\begin{equation} \label{dis}
 \sigma_z(Z)  = 2 \pi \delta(z-z_0)
\end{equation}

At this stage one must take on faith that the field theory expression
is relevant to the supergravity solution. In other words we have assumed
the duality to obtain this result. We will now explore the scalar operators
encoded in the supergravity solution and show that they are consistent
with this distribution function providing a non-trivial cross check of the
duality.

\section{Gauge Theory Operators}

We have written the background on the moduli space of the theory in
coordinates where the gauge coupling takes the explicit form 
expected in the dual field theory. The background involves one other 
function given by (\ref{ans}) in these coordinates. If the theories are truly 
dual we would expect them to be different parametrizations of the same
information \cite{jambab}. 
We should therefore be able to interpret (\ref{ans}) in terms
of field theory quantities. 

The coordinate $W$ transforms under two symmetries. The first is the
dilatation symmetry of the 4d gauge theory,
familiar from the usual $AdS/CFT$ \cite{conjecture}. It
is also present in this case, as can be seen from the way that $W$ enters
the gauge coupling as an energy scale or from the requirement of a 
consistent scaling of the metric, but it is  broken
by a number of parameters. Thus $|W|$ has mass dimension one.
$W$ also transforms under the U(1) symmetry of the 2d plane which corresponds
in the field theory to the U(1) symmetry on the complex scalar. 
So looking in 
(\ref{ans}) we can identify the symmetry properties of the constants and hence
match them to field theory quantities. $\Lambda$ has mass dimension one 
and is a U(1) symmetry invariant - 
it plays the role of the strong coupling scale 
in the field theory as is apparent from its apperance in the coupling
(\ref{eq:couple}). 
$K$ has dimension two and is uncharged under the U(1) symmetry
- it contains two components
which we will shortly show can be written as chargeless 
scalar operators or equivalently as moments of the D5 distribution. 
Finally the function $G$ contains 
a dimension 2 operator
of charge two which we shall again match to a scalar operator.

The ${\cal N} = 2$ field theory on moduli space should be described by
the running coupling and the scalar operators.
We have deduced the distribution function for the vevs $\sigma_z$ 
above (\ref{dis}) 
from the form of the running coupling and hence can calcuate these
functions to see if they match those in (\ref{ans}). There are two dimension 
two operators we can calculate corresponding in the field theory to the
chargeless $tr |\phi |^2$ and the charge two $tr \phi^2$. We must calculate 
these operators in the physical coordinates, $W$.

\beq
{\cal O}_2 = {1 \over 4 \pi} \int \sigma_w(W) W^2 dW d \bar{W} = 
{1 \over 4 \pi} \int \sigma_z(Z) W(Z)^2 dZ d\bar{Z}  = 2 b^2 \Lambda^2
\eeq

\beq
{\cal O}_0 = {1 \over 4 \pi}\int \sigma_w(W) W W^\dagger dW d \bar{W} =
{1 \over 4 \pi}\int \sigma_z(z) |W(Z)|^2  dZ d\bar{Z}  =  
z_0^2 + {b^4 \Lambda^4\over z_0^2}
\eeq

Pleasingly these functional forms precisely match the coefficient of
the gauge coupling in $K$ (\ref{k}) 
and the operator in $G$ (\ref{g}). We are left to explain 
the form of the first term in $K$ which is not one of these moments.
However, it is clear from (\ref{ans}) that the solution contains the 
quantity $\sqrt{ tr \phi^2 tr \phi^{\dagger 2}}$ which is chargeless and
dimension two. This first term can be written as a combination of the 
two chargeless operators. Thus we can write

\begin{equation}
K=\frac{1}{4i}  (\sqrt{   {\cal O}_0^2 - {\cal O}_2 {\cal O}_2^\dagger})
-\frac{1}{2 g^2_{YM}({\rm sing})}{\cal O}_0
\end{equation}
and
\begin{equation}
G=1-\frac{2{\cal O}_2}{W^2}.
\end{equation}

The encoding of the operators in (\ref{ans}) 
is quite complicated but it is encouraging 
that the correspondence can be made between the two duals. 
It is also nice that the distribution function determined above from the
gauge coupling does indeed
match to the functional form of the 
operators parametrized by the rest of the background.
Note that this constitutes the first cross check 
of the assumption in (\ref{ft})
that the coupling of the probe world volume theory in the supergravity 
background is indeed governed by the
field theory expression for the running coupling. 

\section{Conclusion}

We have studied the supergravity solutions found in \cite{martelli, zaff} 
which were 
obtained by studying 7d gauged supergravity and then lifting the solutions
to 10d. The solutions are expected to be the near horizon geometries of D
branes wrapped on S$^2$ and to be dual to ${\cal N} = 2$ SYM theory in 4d.
Following \cite{zaff} we have identified the unique coordinates in
which the theory on the world volume of a probe D5 brane takes ${\cal N} = 2$ 
form. Restricting to the subspace of the background that describes the
field theory's moduli space where these coordinates are known, we have shown
that the background is described by two functions. One of these is the running
gauge coupling of the field theory whilst we have shown the other parametrizes
the field theory operators. Using the field theory expectation for the
form of the running gauge coupling as a function of the D5 distribution,
that distribution can be determined. We have shown that the scalar operators
corresponding to the moments of this distribution function match the form
of the parameters in the second function determining the background. The end 
result is remarkably clean showing that the two dual 
descriptions do indeed encode 
the same physical content, as has been previously observed in ${\cal N} = 4$ 
SYM on moduli space and its gravity dual \cite{brr,freed2, jambab}. 
The result also confirms that
the supergravity background is controlled by the gauge theory dynamics and
that the only renormalization is through the gauge coupling.

Understanding how the gravity background encodes the dual field theory
operators is hopefully a major step towards enlarging the class of known
solutions. In particular the function $G$ in the background (\ref{ans})
looks ripe to be interpreted in general as an 
harmonic function of the two dimensional Laplacian. To confirm whether
such an extension of the solution is possible requires more work than that
presented here since to test a solution of the supergravity
equations one needs more than a restricted subspace of the solution
as we have.  Understanding these backgrounds off the field theory
moduli space, where the field 
theory is less well understood, 
is an important challenge for the future. 

\vskip .2in
\noindent
{\bf Acknowledgements}\vskip .1in
\noindent
The authors thank Michela Petrini, Clifford Johnson, Alex Buchel, Jerome
Gauntlett and Alberto Zaffaroni for discussions.
N.E is grateful for the support of a PPARC Advanced Fellowship
and J.B. for the support of a PPARC Studentship.

\end{document}